# Effect of near-wall treatments on airflow simulations

N. El Gharbi [1,3], R. Absi [2], A. Benzaoui [3] and E.H. Amara [4]

[1] Renewable Energy Development Center, Po. Box 62 Bouzareah 16340 Algiers, Algeria

[2] EBI, Inst. Polytech. St-Louis, Cergy-University, 32 Boulevard du Port, 95094, Cergy-Pontoise Cedex, France

[3] University of Sciences and Technology Houari Boumediene, Po. Box 32 El Alia Bab Ezzouar 16111 Algiers, Algeria

[4] Advanced Technologies Development Centre CDTA/DMIL/TML, Po. Box 17 Baba-Hassen 16303, Algiers, Algeria

**Abstract— Airflow simulation results depend on a good prediction of near wall turbulence. In this paper a comparative study between different near wall treatments is presented. It is applied to two test cases: (1) the first concerns the fully developed plane channel flow (i.e. the flow between two infinitely large plates). Simulation results are compared to direct numerical simulation (DNS) data of Moser et al. (1999) for *Re* = 590 (where *Re* denotes the friction Reynolds number defined by friction velocity *u* , kinematics viscosity  and the channel half-width ); (2) the second case is a benchmark test for room air distribution (Nielsen, 1990). Simulation results are compared to experimental data obtained with laser-doppler anemometry.**

**Simulations were performed with the aid of the commercial CFD code Fluent (2005). Near wall treatments available in Fluent were tested: Standard Wall Functions, Non Equilibrium Wall Function and Enhanced Wall Treatment. In each case, suitable meshes with adequate position for the first near-wall node are needed.**

**Results of near-wall mean streamwise velocity $U^+$ and turbulent kinetic energy $k^+$ profiles are presented, variables with the superscript of + are those non dimensional by the wall friction velocity *u* and the kinematic viscosity .**

*Keywords-component; near wall treatment; airflow; simulation;*

## I. Introduction

Indoor air quality (IAQ) depends greatly on accurate tools for prediction of airflow and dispersion of particles indoors. These particles have potential harmful effects since they may be inhaled by the occupants.

In some work environments, understanding of dispersion and deposition can improve workers safety. In order to provide exposure assessment, numerical simulations are required to allow a better understanding of particles deposition and dispersion indoors.

Reynolds-averaged Navier–Stokes (RANS) turbulent models (such as *k-* models) are still widely used for engineering applications because of their relatively simplicity and robustness. However, these models depend on adequate near-wall treatments.

Airflow simulations depend on a good prediction of near wall turbulence. In our study, different near wall treatments will be assessed and applied to two test cases. The first concerns a fully developed plane channel flow (i.e. the flow between two infinitely large plates), simulations results are compared to direct numerical simulation (DNS) data of Moser et al. (1999) [1] for *Re* = 590 (where *Re* denotes the friction Reynolds number defined by friction velocity *u* , kinematic viscosity  and the channel half-width  ). The second case is a benchmark test for 2D room air distribution (Nielsen, 1990) [2]. The simulation results are compared with experimental data obtained with laser-Doppler anemometry.

All different near wall treatments available in Fluent will be tested: Standard Wall Functions, Non Equilibrium Wall Function and Enhanced Wall Treatment. We will investigate both effect of meshes and position of the first near-wall node.

Simulations will be performed with the aid of the commercial CFD code Fluent (2005) [3]. The *k-* turbulence model, which presents the advantage that it doesn't need excessive computational times, will be used.

## II. Model equations

### A. Governing Equations

Airflow is modeled using the *k-* model. The general form of the governing equations is:

$$\frac{\partial(\rho\phi)}{\partial t} + \text{div}(\rho U\phi) = \text{div}(\Gamma_\phi \text{grad}\phi) + S_\phi \quad (1)$$

Table 1 lists the diffusion coefficients and source terms for the different scalar qualities.

TABLE I
DIFFUSION TERMS AND SOURCE TERMS IN THE GOVERNING EQUATIONS

| Item | Variable | $\Gamma_\phi$ | $S_\phi$ |
|---|---|---|---|
| Continuity | 1 | 0 | 0 |
| x velocity | u | $\mu_{eff} = \mu + \mu_t$ | $-\frac{\partial P}{\partial x} + \frac{\partial}{\partial x}\left(\mu_{eff}\frac{\partial u}{\partial x}\right) + \frac{\partial}{\partial y}\left(\mu_{eff}\frac{\partial v}{\partial x}\right) + \frac{\partial}{\partial z}\left(\mu_{eff}\frac{\partial w}{\partial x}\right)$ |
| y velocity | v | $\mu_{eff} = \mu + \mu_t$ | $-\frac{\partial P}{\partial y} + \frac{\partial}{\partial x}\left(\mu_{eff}\frac{\partial u}{\partial y}\right) + \frac{\partial}{\partial y}\left(\mu_{eff}\frac{\partial v}{\partial y}\right) + \frac{\partial}{\partial z}\left(\mu_{eff}\frac{\partial w}{\partial y}\right) - \rho g$ |
| z velocity | w | $\mu_{eff} = \mu + \mu_t$ | $-\frac{\partial P}{\partial z} + \frac{\partial}{\partial x}\left(\mu_{eff}\frac{\partial u}{\partial z}\right) + \frac{\partial}{\partial y}\left(\mu_{eff}\frac{\partial v}{\partial z}\right) + \frac{\partial}{\partial z}\left(\mu_{eff}\frac{\partial w}{\partial z}\right)$ |
| Kinetic energy | k | $\alpha_k \mu_{eff}$ | $G_k + G_b - \rho\varepsilon$ |
| Dissipation rate | $\varepsilon$ | $\alpha_\varepsilon \mu_{eff}$ | $C_{1\varepsilon}\frac{\varepsilon}{k}(G_k + C_{3\varepsilon}G_b) - C_{2\varepsilon}\rho\frac{\varepsilon^2}{k} - R_\varepsilon$ |
| Temperature | T | $\frac{\mu}{Pr} + \frac{\mu_t}{\sigma_T}$ | $S_T$ |
| Concentration | C | $\frac{\mu}{Sc} + \frac{\mu_t}{\sigma_c}$ | $S_c$ |

### B. Near-wall treatments

#### 1) Standard Wall Functions

The standard wall functions in Fluent are based on the proposal of Launder and Spalding (1974) [4], and have been most widely used for industrial flows.

#### 2) Non Equilibrium Wall Function

Kim and Choudhury (1995) [5] proposed the use of the Non Equilibrium Wall Function in order to improve the accuracy of the standard wall functions. Because of the capability to partly account for the effects of pressure gradients and departure from equilibrium, the non-equilibrium wall functions are recommended for use in complex flows involving separation, reattachment, and impingement where the mean flow and turbulence are subjected to severe pressure gradients and change rapidly [3].

For these two wall functions, the first cell must be in Log Layer region.

#### 3) Enhanced Wall Treatment

Enhanced wall treatment is a near-wall modeling method that combines a two-layer model with enhanced wall functions. Fine meshes: two-layer approach (Wolfstein, 1969 [6], Chen and Patel, 1988 [7]) and coarse meshes: enhanced wall-function approach (Kader, 1993 [8]).

#### 4) Analytical near-wall TKE profile

Absi (2008) [9] suggested a general equation for the turbulent kinetic energy $k^+$ in the near-wall region (for $y^+$ 20) as:

$$k^+ = B(y^+)^2 e^{\left[-\frac{y^+}{s}\right]} \qquad (2)$$

B is a coefficient which depends on Re (Absi, 2009 [10]).

### III. TEST CASES

Airflow simulations with different near-wall treatments are applied to two test cases:

#### A. Channel flow

The first test case is the fully developed plane channel flow (i.e. the flow between two infinitely large plates, figure 1). Simulations results are validated by direct numerical simulation (DNS) data of Moser et al. (1999) [1] for $Re = 590$.

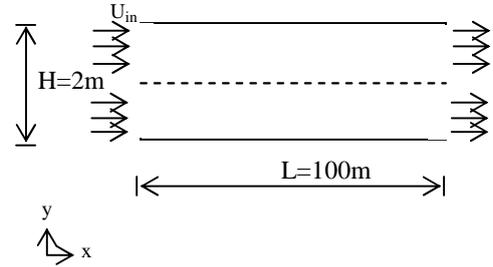

Figure 1. presentation of the channel flow

#### B. Room air distribution

The second test case is a benchmark test for a room air distribution (Nielsen, 1990 [2], figure 2). The simulation results are validated by experimental data obtained with laser-doppler anemometry.

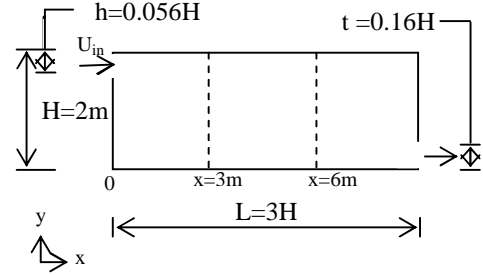

Figure 2. Presentation of Nielsen room, H=3m and L=9m.

### IV. RESULTS AND DISCUSSIONS

All different near wall treatments available in Fluent were tested: Standard wall functions "SWF", Non equilibrium wall function "NEWF" and Enhanced wall treatment "EWT".

Results of mean streamwise velocity $u^+$ and turbulent kinetic energy $k^+$ profiles are presented in figures (3) and (6).

For the two test cases, channel flow and room air distribution, a fine mesh (respectively 500×57 and 45×38) was used for enhanced wall treatment "EWT", while a coarse mesh (respectively 500×19 and 45×12) was used for standard wall function "SWF" and non-equilibrium wall function "NEWF" (figure 4).

For the first test case (fully developed plane channel flow), figure 3 presents simulation results: mean streamwise velocity $u+$ (fig. 3.a) and turbulent kinetic energy "TKE" $k^+$ (fig. 3.b) profiles, with DNS data of Moser et al. (1999) [1] for $Re = 590$.

On the one hand, standard "SWF" and non equilibrium "NEWF" wall functions need a coarse mesh (fig. 4.a). The first node should be at $y^+>30$. Figure (3) shows that standard "SWF" and Non equilibrium "NEWF" wall functions predict well velocity profiles for $y^+>30$ and "TKE" profiles for $y^+>60$. However, these near wall treatments are not able to provide details about velocity and TKE in the viscous and buffer layers. If these treatments are used, it is possible to provide an accurate description of TKE (figure 5, solid line) by equation (2) (Absi, 2008) and velocity by solving an ordinary differential equation "ODE" (Absi, 2009). These treatments could be therefore associated to this simple and efficient analytical method.

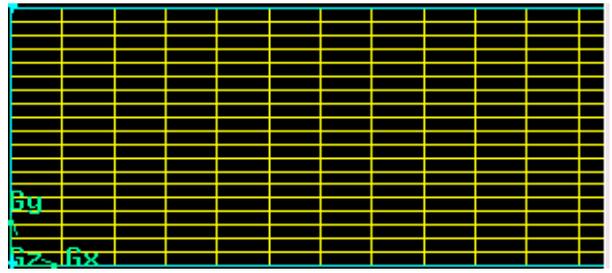

(a)

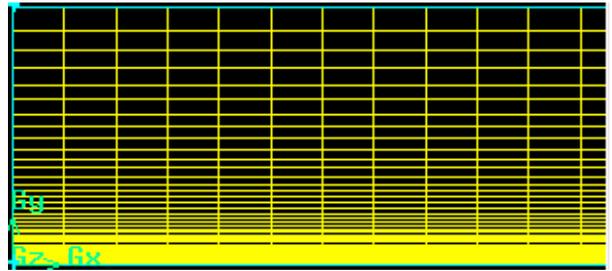

(b)

Figure 4. Used meshes; (a) for standard and non-equilibrium wall functions, (b) for enhanced wall treatments

On the other hand, enhanced wall treatment "EWT" needs a finest mesh in the viscous sublayer (fig. 4.b). The first node should be at about $y^+=1$. Figure (3) shows that the velocity profile is more accurate and well predicted even in the viscous and buffer layers. However, TKE is underestimated (fig. 3.b). This has no effect on velocity profile but can provide an underestimated eddy viscosity/diffusivity which could be involved in predicted particles concentrations.

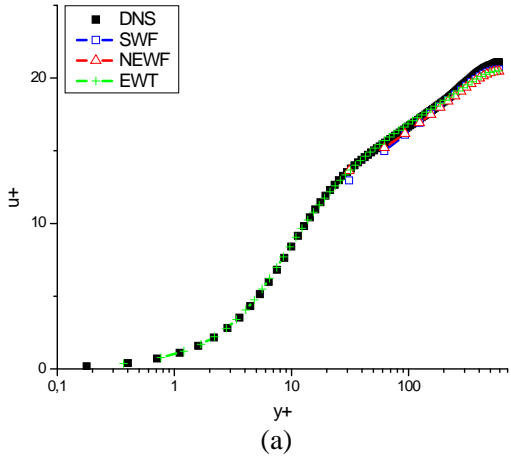

(a)

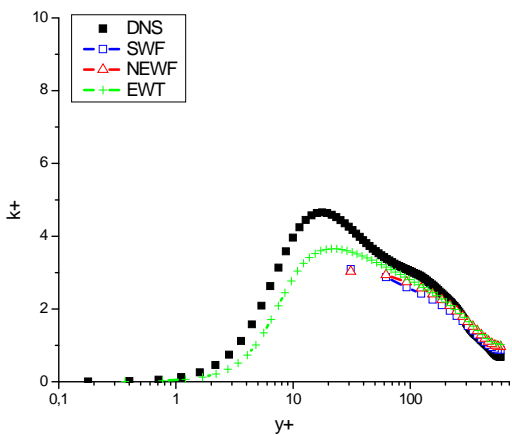

(b)

Figure 3. Comparison between predicted profiles using standard $k$-model with different wall treatments and DNS data for test case 1 fully diveloped plane channel flow. (a) mean stremwise velocity, (b) turbulent kinetic energy

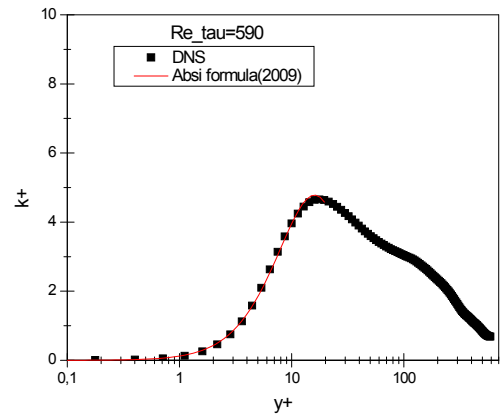

Figure 5. Comparison between predicted TKE by equation (2) and DNS data

In order to investigate the effet of standard $k$- model on the TKE profile wich is underestimated by "EWT" (fig. 3.b), figure (6) presents a comparison with Re-Normalisation

Group "RNG" $k$-$\varepsilon$ model. Figure (6) shows that RNG $k$-$\varepsilon$ model provides a very small improvement for velocity and TKE. Since the difference is negligeable, the underestimation of TKE seems therefore not related to the used turbulence model but associated to the near wall treatment.

Predicted mean velocity profiles with the different near-wall treatments are quite similar (fig. 7.a, 7.c). Mean velocities obtained with enhanced wall treatment "EWT" seem better particularly near the walls where wall functions are unable to provide values. However, EWT needs more computation time.

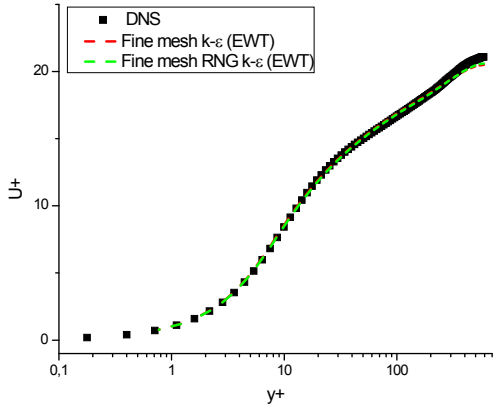

(a)

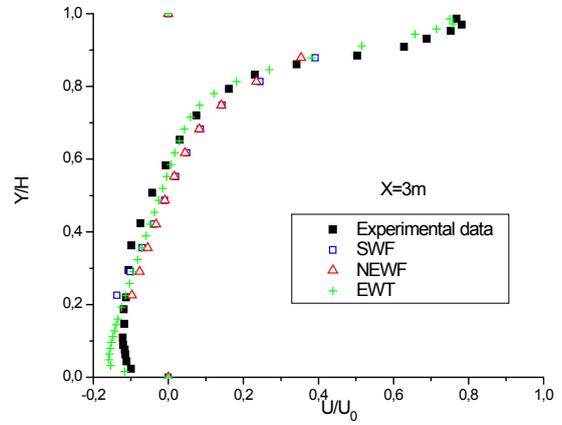

(a)

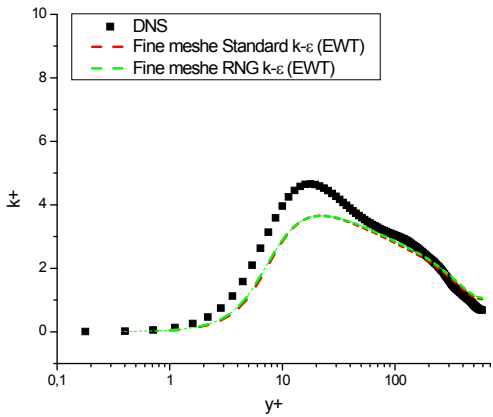

(b)

Figure 6. Comparison between predicted profiles using standard and RNG $k$-$\varepsilon$ models with enhanced wall treatmant "EWT" and DNS data for test case 1 fully divelloped plane channel flow. (a) mean stremwise velocity, (b) turbulent kinetic energy

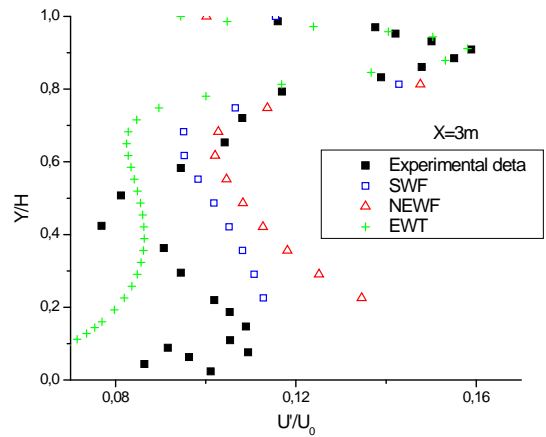

(b)

In order to improve TKE, we suggest the use of equation (2) for $y^+ \leq 20$. The value of TKE at $y^+=20$ could be used as a boundary condition for the modeled k-equation for $y^+>20$. Since TKE is well predicted until 20 by Eq. (2), the improvment of TKE for $y^+>20$ is expected.

The second test case (benchmark test for a room air distribution), presents simulation results: mean velocity $u^+$ (fig. 7a and 7.c) and turbulence intensity (figure 7.b and 7.c), with experimental data obtained by laser-Doppler anemometry (Nielsen, 1990) [2].

Figures (7.a) and (7.c) present mean velocity $u^+$ respectively at $x$=3m (1/3 L) and $x$=6m (2/3 L) while figures (7.b) and (7.d) present turbulence intensity $u'$ (respectively at $x$=3m and $x$=6m).

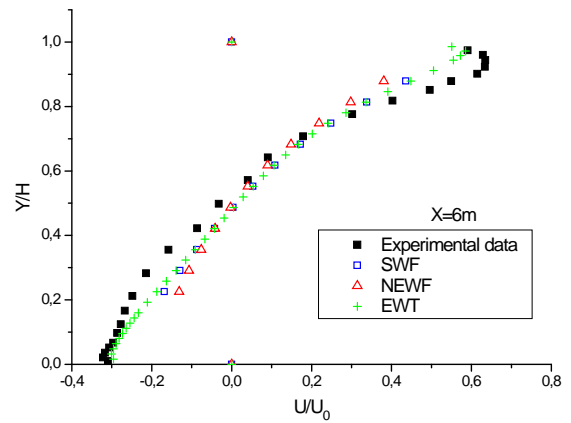

(c)

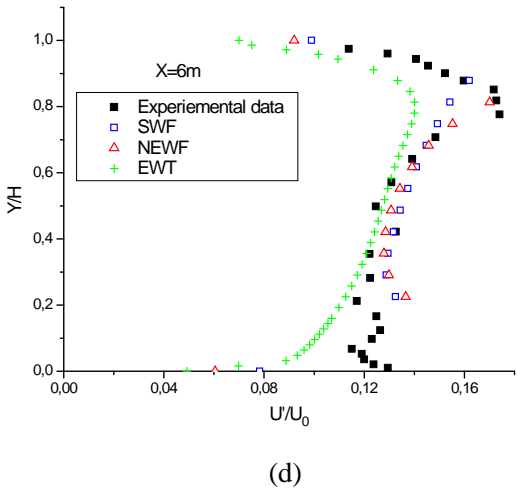

(d)

Figure 7. Comparison between predicted profiles using standard k-model with different wall treatments and experimental data for test case 2 benchmark test for a room air distribution. (a) mean velocity at *x*=3m, (b) RMS velocity at *x*=3m, (c) mean velocity at *x*=6m, (d) RMS velocity at *x*=6m.

More important scatter is shown for RMS (root mean square) velocities at *x*=3m (fig. 7.b). Non equilibrium wall function seems to be the less accurate. All near-wall treatments fail to predict RMS velocities for *y/H*<0.4 (fig. 7.b). In contrast, at *x*=6m wall functions seem more accurate for *y/H*>0.6. However, for *y/H*<0.2 wall functions (SWF and NEWF) didn't provide values, this is due to the required mesh and first near wall node, while EWT seems not accurate in this region.

## V. CONCLUSIONS

Airflow simulations with different near-wall treatments were applied to two test cases.

For the first test case (fully developed plane channel flow), simulation results: mean streamwise velocity and turbulent kinetic energy "TKE" profiles were compared to DNS data for *Re* = 590. Standard "SWF" and non equilibrium "NEWF" wall functions need a coarse mesh. The first node should be at $y^+>30$. "SWF" and "NEWF" wall functions predict well velocity profiles for $y^+>30$ and "TKE" profiles for $y^+>60$. But they are not able to provide details about velocity and TKE in the viscous and buffer layers. It is possible to provide an accurate description of TKE by equation (2) (Absi, 2008) and velocity by solving an ordinary differential equation (Absi, 2009). Enhanced wall treatment "EWT" needs a finest mesh in the viscous sublayer. The first node should be at about $y^+=1$. Velocity profile is more accurate and well predicted even in the viscous and buffer layers. TKE is underestimated which could provide an underestimated eddy viscosity/diffusivity and therefore could had an effect on predicted particles concentrations. Simulations show no difference between standard and RNG k- models. The underestimated TKE seems therefore associated to near wall treatments. In order to improve TKE, we suggest the use of equation (2) (Absi, 2008) for $y^+$ 20.

The value of TKE at $y^+=20$ could be used as a boundary condition for the modeled k-equation for $y^+>20$.

For the second test case (benchmark test for a room air distribution) simulation results for mean velocity and turbulence intensity (at *x/L*=1/3 and 2/3) were compared to experimental data. Predicted mean velocity profiles with the different near-wall treatments are quite similar. Mean velocities obtained with enhanced wall treatment "EWT" seem better particularly near the walls. However, "EWT" needs more computation time. More important scatter is shown for RMS velocities at *x/L*=1/3. Non equilibrium wall function seems to be the less accurate. All near-wall treatments fail to predict RMS velocities for *y/H*<0.4. In contrast, at *x/L*=2/3 wall functions seem more accurate for *y/H*>0.6. However, for *y/H*<0.2 no values are obtained by wall functions (SWF and NEWF), this is due to the required mesh and first near wall node, while "EWT" seems not accurate in this region. Improved models with adequate near-wall treatments are needed for an efficient simulation of room air distribution.